\documentstyle[11pt,newpasp,twoside,epsf]{article}
\def\chandra    {{\em Chandra}\/}
\def\xmm        {{\em XMM}\/}
\def\rosat      {{\em ROSAT}\/}

\def\as         {$^{\prime\prime}$}

\def\lesssim{\mathrel{\hbox{\rlap{\hbox{\lower4pt\hbox{$\sim$}}}\hbox{$<$}}}}
\def\gtrsim{\mathrel{\hbox{\rlap{\hbox{\lower4pt\hbox{$\sim$}}}\hbox{$>$}}}}
\def\lax{\lesssim}

\def\hfifty     {$H_0$=50~km$\;$s$^{-1}\,$Mpc$^{-1}$}

\begin{document}

\title{A high resolution picture of the intracluster gas}

\author{M. Markevitch, A. Vikhlinin, W. R. Forman}

\affil{Smithsonian Astrophysical Observatory, Cambridge, MA 02138, USA}

\begin{abstract}
\chandra\ has significantly advanced our knowledge of the processes in the
intracluster gas.  The discovery of remarkably regular ``cold fronts'', or
contact discontinuities, in merging clusters showed that gas dynamic
instabilities at the boundaries between the moving gases are often
suppressed, most likely by specially structured magnetic fields.

Cold fronts are not limited to mergers --- \chandra\ observed them in the
cores of more than 2/3 of the cooling flow clusters, where they often divide
the cool central and the hotter ambient gas phases.  The natural state of
the low-entropy central gas in clusters thus appears to be to slosh
subsonically in the central potential well, with several interesting
implications.  Hydrostatic equilibrium is not reached and the hydrostatic
total mass estimates cannot be valid at small radii ($r\lax 100$ kpc).  The
kinetic energy dissipated by the sloshing gas may be sufficient to
compensate for radiative cooling.  In the absence of a recent merger, the
cause of such sloshing is unclear.  It might be related to the mechanical
energy generated by the central AGN, evidence of which (the bubbles) is also
observed by \chandra\ in a large fraction of the cooling flow clusters.

Heat conduction across cold fronts is suppressed completely.  In the bulk of
the gas, it is reduced by a factor of $\sim 10$ or more relative to the
classic value, as shown by merger temperature maps.  A spatial correlation
between the gas temperature maps and the radio halo maps is observed in
several mergers, which may provide clues for the relativistic particle
acceleration mechanisms.
\end{abstract}

\section{Introduction}

\begin{figure}
\plotone{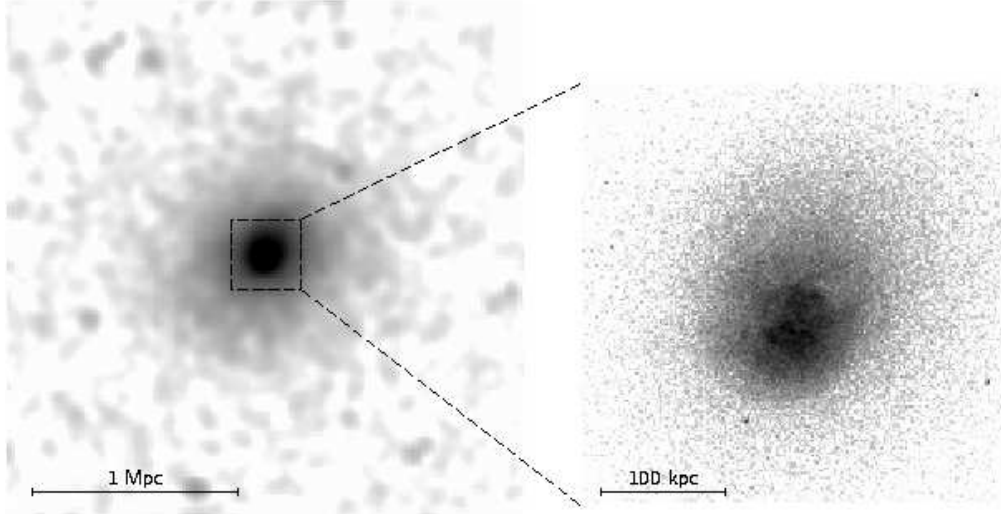}
\caption{{\em Left}: \rosat\ PSPC image of the relaxed cluster 2A0335+096
  ($z$=0.03, $T$=3 keV). {\em Right}: \chandra\ image reveals complex and
  dynamic structure of its core.}
\label{fig:2a0335}
\end{figure}

\chandra\ X-ray Observatory's salient feature is its 1\as\ resolution.  The
difference it makes for the galaxy cluster studies is illustrated in Fig.\
\ref{fig:2a0335}, which shows images of one of the most relaxed nearby
clusters, 2A0335+096, made by \rosat\ PSPC and \chandra.  On large linear
scales accessible to \rosat, the cluster is symmetric and, apparently, in
hydrostatic equilibrium.  However, in the central 100 kpc, \chandra\ reveals
a very dynamic gas core.  In this article, we will review the published
results and report some new findings on dynamic phenomena in the cluster
gas.  Some other interesting \chandra\ work, such as the accurate total mass
profiles and studies of the AGN interaction with the cluster gas, is
presented elsewhere in these proceedings.  We use \hfifty.

\section{Gas motion in merging clusters}
\label{sec:merg}

\begin{figure}
\plotone{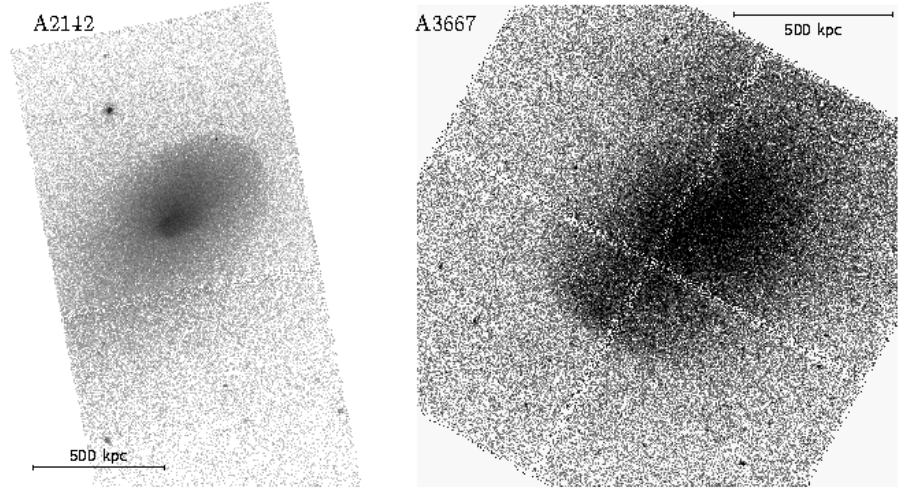}
\caption{\chandra\ images of merging clusters A2142 and A3667 reveal
  prominent gas density edges.}
\label{fig:fronts}
\end{figure}

\begin{figure}[b]
\plotone{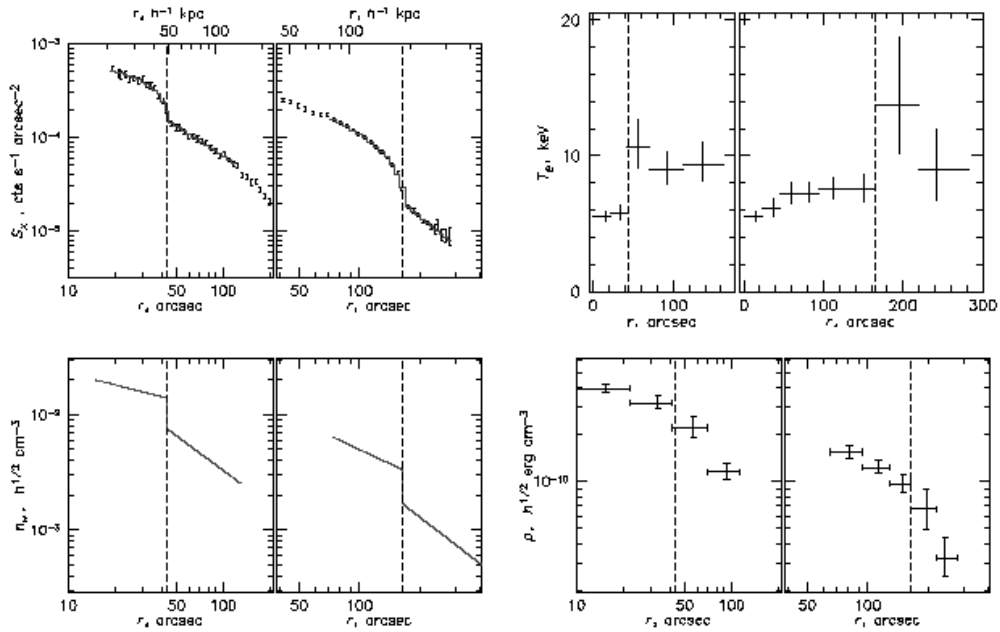}
\caption{{\em Upper panels}: radial profiles of X-ray brightness and
  temperature for A2142 derived in the sectors of two brightness edges (see
  Fig.\ \ref{fig:fronts}). {\em Lower panels}: density model fits to the
  brightness profiles (their projections are overlaid on the brightness
  profiles as histograms) and gas pressure profiles.  Density and
  temperature jump while pressure stays nearly continuous across these and
  other ``cold fronts'' (from Markevitch et al.\ 2000).}
\label{fig:2142prof}
\end{figure}

\begin{figure}[t]
\plotone{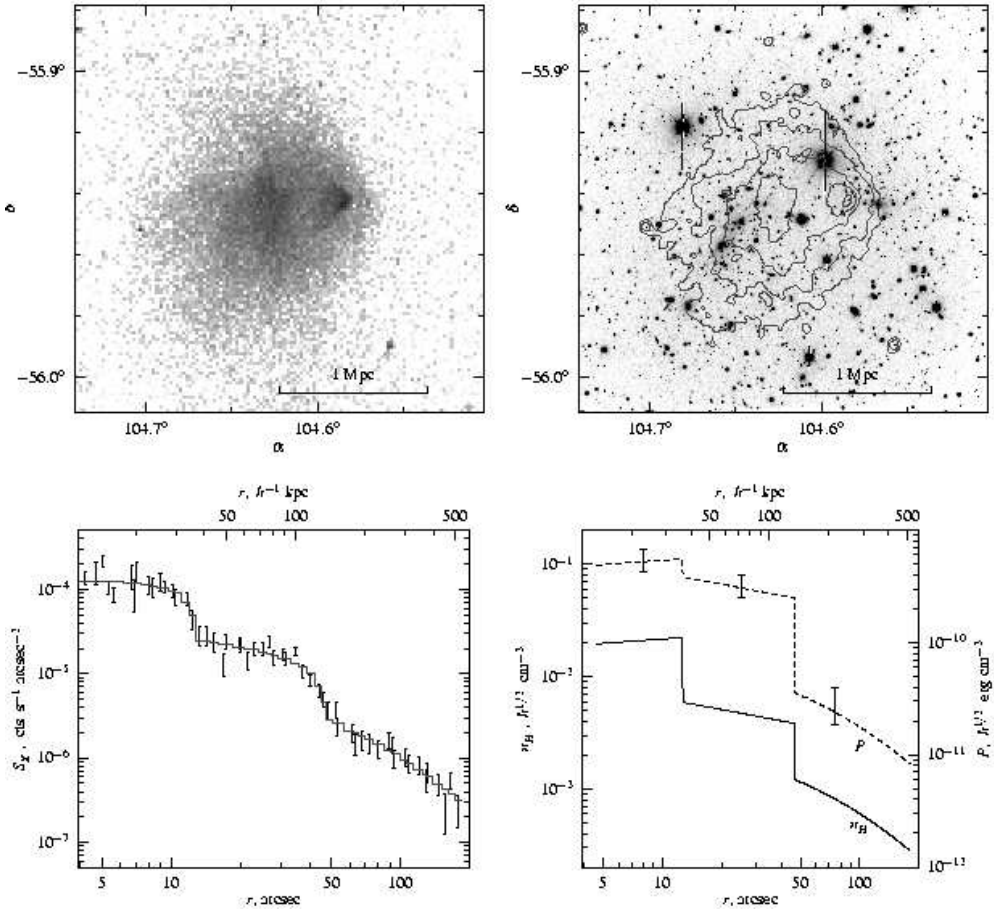}  
\vspace{-3mm}
\caption{A merging cluster 1E0657--56 ($z$=0.3, $T$=15 keV): \chandra\
  X-ray image; its overlay (as contours) on the optical plate (courtesy of
  E. Falco and M. Ramella); X-ray brightness profile centered on the
  ``bullet'' and derived in the western sector containing the two density
  edges; a gas density model in that sector, and the pressure profile. The
  first edge is a cold front while the second is a bow shock (from
  Markevitch et al.\ 2002).}
\label{fig:1e}
\end{figure}

Among the first \chandra\ cluster results was a discovery of ``cold fronts''
in merging clusters A2142 and A3667 (Markevitch et al.\ 2000; Vikhlinin,
Markevitch, \& Murray 2001ab).  Figure \ref{fig:fronts} shows ACIS images of
A2142 and A3667 with prominent X-ray brightness edges.  A close examination
(Fig.\ \ref{fig:2142prof}) shows that (a) these edges are perfectly modeled
by an abrupt, spherically symmetric (within a certain sector) jump of the
gas density, (b) the gas on the denser side of the edge is cooler, the
opposite to what is expected for a shock front, and (c) there is approximate
pressure equilibrium across the edge, or at least no large pressure jump
such as expected in a shock front.  We proposed that these edges are contact
discontinuities, or ``cold fronts'', delineating the boundaries of the dense
cool cores of merging subclusters that have survived shocks and destruction
of a merger and are now moving, along with the underlying dark matter
clumps, through a shock-heated surrounding gas.  A more recent observation
of another merging cluster, 1E0657--56 (Fig.\ \ref{fig:1e}), illustrates
this quite clearly (Markevitch et al.\ 2002).  It reveals a cool gas
``bullet'' moving westward, that lags behind its underlying galaxy
subcluster seen in the optical plate, because of ram pressure.  The bullet's
sharp front edge is a cold front; further ahead in front of the bullet,
there is a gas density jump of a similar amplitude which is a genuine bow
shock.  Cold fronts have since been observed in several other mergers (e.g.,
Sun et al.\ 2001; Kempner et al.\ 2002) and reproduced in simulations (e.g.,
Nagai \& Kravtsov 2002; Bialek et al.\ 2002), although the spatial
resolution of the present simulations is far behind that of the X-ray data.
Shock fronts are less easily observed, and only one other than the shock in
1E0657 has been unambiguously detected so far, a weak one in A3667.

Cold fronts and shocks offer unique insights into the cluster physics,
including determining the gas bulk velocity, its acceleration, growth of
plasma instabilities, strength and structure of magnetic fields, and thermal
conductivity.  From the pressure profile across the cold front in A3667,
Vikhlinin et al.\ (2001b) determined the Mach number of the subcluster to be
$1\pm0.2$.  The temperature and density jumps at the shock in 1E0657 give
the Mach number of the ``bullet'' as $2-3$ (Markevitch et al.\ 2002).

The extent and regularity of the observed cold fronts
(Fig.\ \ref{fig:fronts}) is unexpected.  Vikhlinin et al.\ (2001a) estimated
that, given the measured subcluster velocity, the front in A3667 should have
been destroyed by Kelvin-Helmholtz instability.  Its existence thus requires
partial suppression of the instabilities, for example, by surface tension of
a layer of the magnetic field parallel to the front, with a field strength
of $\sim 10\;\mu$G.  Such a field configuration may be created by the
tangential gas flow.  Once formed, it would effectively stop the plasma
diffusion and heat conduction across the front (as is indeed observed, see
\S 4), and may inhibit gas mixing during the subcluster merger.

The gas bullet in 1E0657 (Fig.\ \ref{fig:1e}) is near the final stage of
being destroyed by ram pressure and gas dynamic instabilities.  A longer
\chandra\ exposure has just been obtained and will show these processes in
exquisite detail.

The cold fronts discussed above are observed in merging clusters, with
orbiting subclusters driving the gas motion by their gravitational pull.  In
A3667, the X-ray surface brightness distribution inside the cold front even
allows a determination of the precise center of the underlying dark matter
clump (inside the front but well ahead of the cool gas centroid; Vikhlinin
\& Markevitch 2002).  There is a different class of cold fronts observed in
clusters with no signs of merging; it is discussed below.

\section{Non-hydrostatic gas in cores of relaxed clusters}

\begin{figure}[t]
\plotone{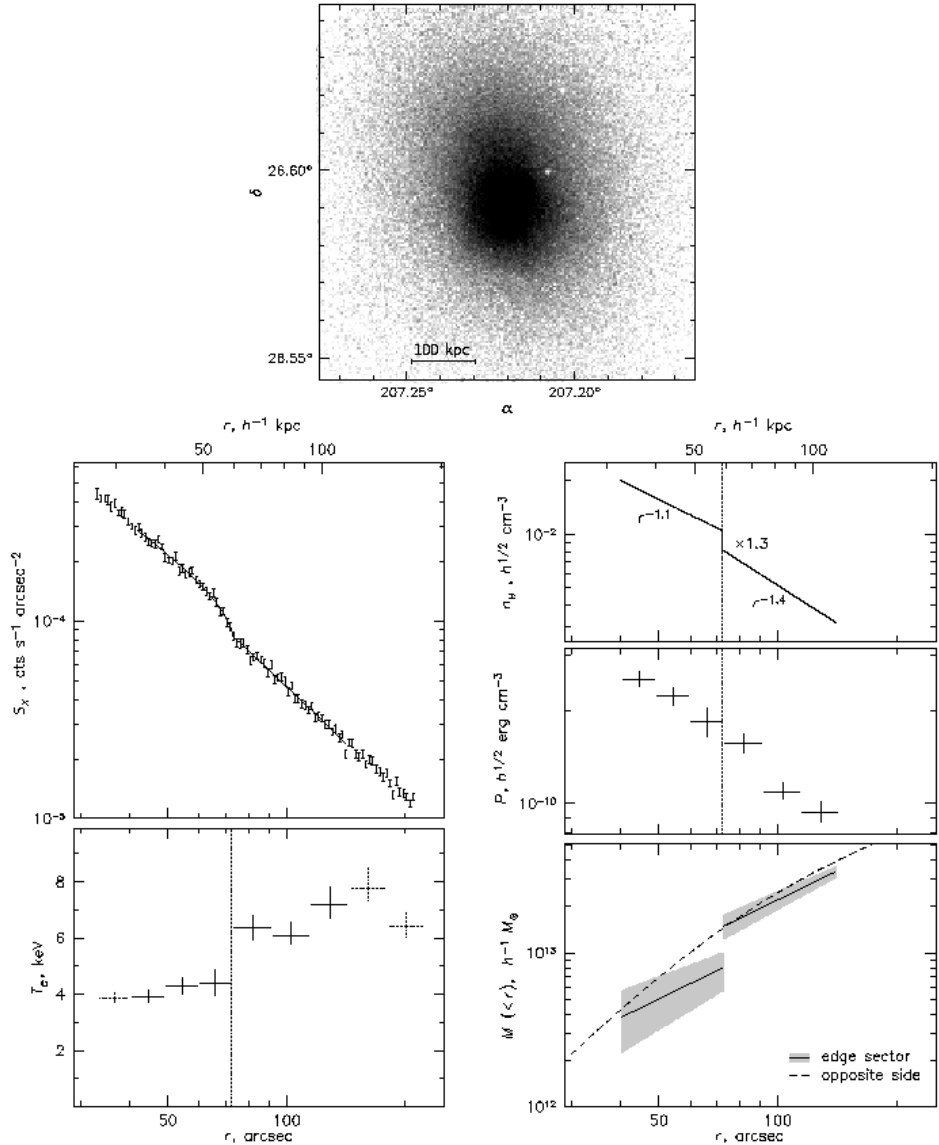}
\vspace{-4mm}
\caption{A relaxed cluster A1795 ($z$=0.06, $T$=7 keV): its \chandra\ image
  and profiles of X-ray brightness and temperature ({\em left}) in the
  southern sector that contains a subtle brightness edge. {\em Right}\/
  panels show a density model with a jump, a continuous pressure profile,
  and a total mass profile derived from these data under the (wrong)
  hydrostatic equilibrium assumption. A mass profile derived from the
  opposite sector is shown for comparison (from Markevitch et al.\ 2001).}
\label{fig:a1795}
\end{figure}

Cold fronts similar to those in A2142 and A3667, but much less prominent,
were reported in the cores of relaxed clusters RXJ1720.1 (Mazzotta et al.\
2001) and A1795 (Markevitch, Vikhlinin \& Mazzotta 2001).  The latter is
shown in Fig.\ \ref{fig:a1795} (a subtle brightness edge in the southern
sector) along with the radial profiles of the gas density, temperature and
pressure.  This small density jump has interesting implications which we
review here in detail.  The pressure is continuous across the front,
indicating that the current relative velocity of the gases is near zero,
making the edge appear to be in hydrostatic equilibrium.  However, a total
mass profile derived, under the equilibrium assumption, from the radial
profiles of the gas density and temperature in this sector, exhibits an
unphysical jump by a factor of $\sim 2$.  For comparison, we also derived a
mass profile from the data at the opposite, smooth side of the cluster.
Outside the radius of the edge, the two profiles are in agreement, as they
must be if the gas on both sides traces the same centrally symmetric
potential.  Such an apparent mass discontinuity was first reported in
RXJ1720.1 (Mazzotta et al.\ 2001).

We proposed that the cool central gas is ``sloshing'' in the cluster
gravitational potential well and is now near the point of maximum
displacement, where it has zero velocity but nonzero centripetal
acceleration. The distribution of this non-hydrostatic gas should reflect
the reduced gravity force in the accelerating reference frame, resulting in
the apparent mass discontinuity.  Assuming that the gas outside the edge is
hydrostatic, the acceleration of the moving gas can be estimated from the
mass jump $\Delta M$ as $a\sim G\, \Delta M\, r^{-2} \approx 400\;{\rm
km}\;{\rm s}^{-1}\;(10^8\;{\rm yr})^{-1}$, where $r$ is the radius of the
edge.  There is a cool gas filament at the position of the cD galaxy (Fabian
et al.\ 2001), aligned with the apparent direction of the gas motion.  If
the galaxy is at rest in the cluster potential while the gas is flowing
around it, then the length of the filament, $60-80$ kpc, shows the amplitude
of the gas sloshing.

The cold front in A1795 is a boundary between the cool (cooled in situ, or
fallen in) gas in the core and the ambient cluster gas; apparently, they do
not mix readily when stirred, similarly to the gases belonging to different
subclusters in the mergers A2142 and A3667.  Gas motion in the centers of
relaxed clusters has other interesting implications which will be discussed
below
(\S\S\ 3.2-3.3), but first we will check how widespread this phenomenon is.

\subsection{Prevalence of gas sloshing}

\begin{figure}
\centering \leavevmode \epsfxsize=12cm 
\epsfbox{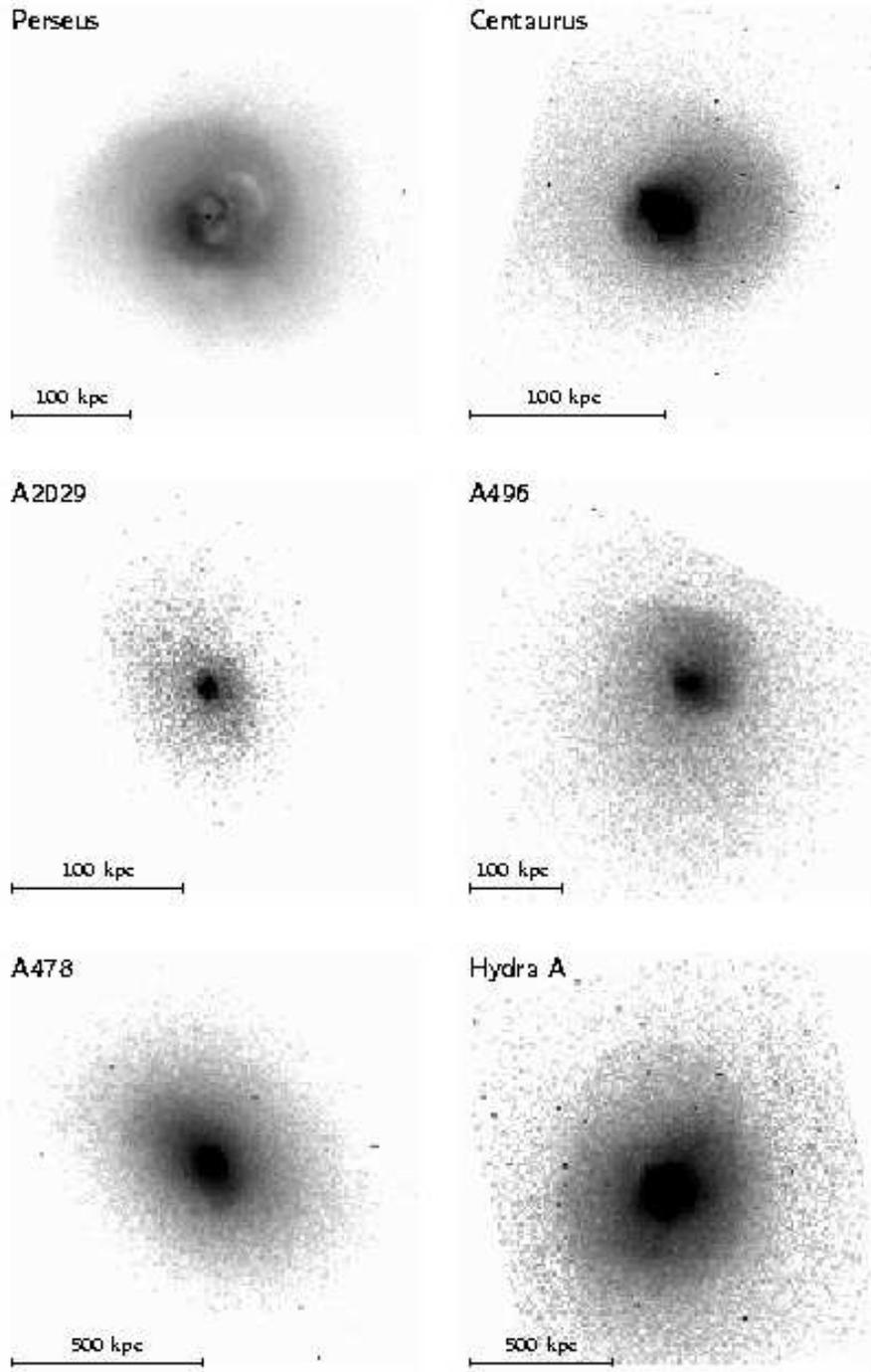}
\caption{\chandra\ images of several examples of cold fronts in
  the cores of relaxed clusters (see Figs.\ 1 and \ref{fig:a1795} for more).
  In some clusters, more than one edge is present.  Gas sloshing appears to
  be a natural state of the cool central gas in clusters.}
\label{fig:gallery}
\end{figure}

We have selected 37 nearby clusters known from earlier X-ray studies to be
symmetric and ``relaxed'', that have archival \chandra\ data of good
statistical quality.  Although the sample is not objective, for our
qualitative purposes it is sufficiently representative of ``relaxed''
clusters.  \chandra\ images for 25 of these clusters (or 2/3) exhibit one or
more X-ray brightness edges in their cores similar to or more prominent than
that in A1795.  Of the 37 clusters, 18 are in the Peres et al.\ (1998)
\rosat\ list of cooling flows; among those, 14 (or 80\%) have edges.  Given
the projection, such frequency implies that cold fronts should be present in
the cores of {\em most if not all}\/ ``relaxed'' clusters.  Several
particularly interesting examples are shown in Fig.\ \ref{fig:gallery} and
Fig.\ \ref{fig:2a0335} (more details will be given in Markevitch et al., in
preparation).

\subsection{Gas sloshing and the total mass estimates}
\label{sec:sloshmass}

Most cold fronts in cooling flow clusters are within $r\sim 100$ kpc of the
cluster center (although some, e.g., in Hydra A, are at greater distances).
The gas at these small radii is not in hydrostatic equilibrium, and if one
makes this wrong assumption and calculates a total enclosed mass, one would
obtain an underestimate, as shown above for A1795.  This should be the case
for most clusters.  It is at these radii where for many clusters, strong
lensing indicates a 2--3 times higher mass than X-ray estimates (e.g.,
Miralda-Escud\'e \& Babul 1995 and later works).  Note, however, that the
volume inside the radii of these edges typically contains only $\sim 1$\% of
the gas within the virial radius, so this finding cannot be generalized to
the whole cluster; indeed, at greater radii, there is an agreement between
the lensing and X-ray masses when accurate gas temperature and density
profiles are used (Allen, Schmitt, \& Fabian 2002 and references therein;
these proceedings).

\subsection{Gas sloshing and cooling flows}
\label{sec:sloshcf}

Another interesting implication is related to the energetics of the cooling
flows.  \xmm\ grating results (Peterson et al.\ 2001 and later works) showed
unambiguously that the amount of cool gas in the cooling flow regions is
much less than expected in the simple cooling flow models, implying that
some steady source of heat balances radiative cooling.  A number of
explanations was proposed (e.g., David et al.\ 2001; Nulsen et al.\ 2001;
Fabian et al.\ 2002ac; Churazov et al.\ 2002).  Gas sloshing may provide
just such a source.  As the gas moves, it dissipates kinetic energy into
heat.  An estimate for A1795 showed that the available potential energy of
the moving gas is comparable to its thermal energy, thus, in principle, such
sloshing can offset radiative cooling (Markevitch et al.\ 2001).  For the
Perseus cluster, one can make a more detailed estimate, comparing the gas
gravitational potential energy $\Delta E$ in the present non-equilibrium
configuration (Fig.\ \ref{fig:gallery}) w.r.t.\ the equilibrium distribution
that the same gas would have, for example, when it settles down in the NFW
potential.  This energy should dissipate into heat on several free fall
timescales, $t$.  An order of magnitude estimate shows that $\Delta E/t \gg
L_X$, where $L_X$ is the X-ray luminosity --- or the radiative cooling rate
--- of the same central region of the cluster (Vikhlinin \& Markevitch, in
preparation).  In a steady state process (if something continually stirs the
gas), the present gas kinetic energy is of order $\Delta E$, so there is
more than enough heat being generated by the sloshing gas to compensate for
radiative cooling.

The original source of this kinetic energy is a separate issue, but in
Perseus, the observed asymmetric gas distribution is probably the result of
the central AGN activity producing hot bubbles that are so prominent in the
\chandra\ and even \rosat\ HRI images (B{\"o}hringer et al.\ 1993; Fabian et
al.\ 2002b and references therein).  The AGN was proposed as the source of
energy to compensate for cooling (e.g., Churazov et al.\ 2000, 2002; David
et al.\ 2001; B{\"o}hringer et al.\ 2002).  But it is an intermittent
source, and the kinetic energy of the gas bulk motions may provide the
necessary intermediate reservoir to release it more steadily.

\subsection{Origin of gas sloshing}

\begin{figure}[b]
\centering \leavevmode \epsfxsize=8cm 
\epsfbox{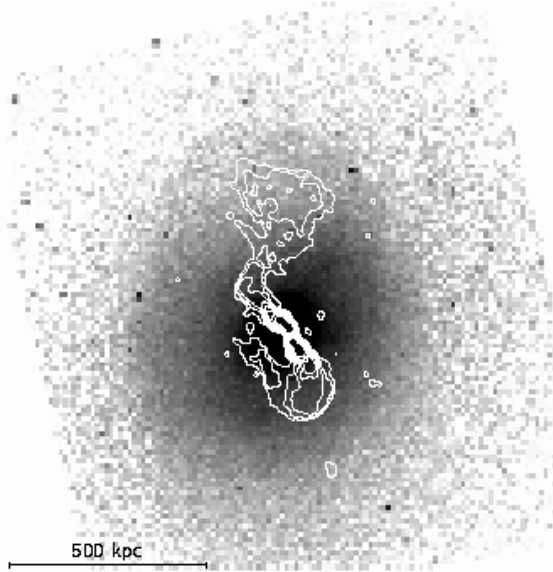}
\caption{\chandra\ image of Hydra A (also shown in Fig.\ \ref{fig:gallery})
  with 90 cm radio contours (courtesy G. Taylor).  The inner, brighter radio
  lobes in the overexposed area of the X-ray image coincide with the X-ray
  cavities discovered by McNamara et al.\ (2000). The northern extended
  radio lobe is just inside the cold front, as in a number of other
  clusters, suggesting a possible causal relation.}
\label{fig:hydrad}
\end{figure}

The cluster gas is expected to have small residual bulk velocity (streams
and turbulence) from a recent merger.  However, many of the clusters
considered in this section do not show any traces of recent mergers --- in
fact, A2029 and A1795 have very symmetric gas halos on scales $r>100-200$
kpc and are the most relaxed among the nearby clusters according to some
measures (e.g., Buote \& Tsai 1996).  Unlike the cold fronts in such mergers
as A3667 and 1E0657 where the moving gas clouds have obvious underlying
moving dark matter clumps, the gas in these clusters apparently sloshes by
itself.  Perhaps a localized gas motion in the core might be caused by a
disturbance of the central gravitational potential by an infalling small
subcluster. Such a disturbance would supply kinetic energy to the
low-entropy gas accumulated over time in the cluster center, which is more
responsive to small potential perturbations than the hotter outer gas.  In
the very center ($r\rightarrow 0$), the gravitational potential of the NFW
radial mass profile has a beak of the form $\varphi \propto r$, compared to
$\varphi = {\rm const}$ for a profile with a core.  If both the cluster and
the infalling subcluster have such cuspy central profiles, it may be
possible to create a localized disturbance that would last sufficiently
long; simulations are needed to study this possibility.

Another possible origin of the gas sloshing is hinted at by the X-ray
cavities and bubbles observed in many cooling flow clusters (for a review,
see McNamara 2002).  In many clusters, these cavities coincide with radio
lobes from the central AGN, and where they don't, a more sensitive study
reveals fainter radio lobes at the locations of the cavities (e.g., Fabian
et al.\ 2002b).  The gas is evacuated from the cavities by expanding ejecta
from the AGN; the resulting bubbles then rise buoyantly and expand (e.g.,
Churazov et al.\ 2001).  In a number of clusters, these bubbles are located
just inside cold fronts.  An example is Hydra A, whose central, bright radio
lobes coincide with the X-ray cavities discovered by McNamara et al.\
(2000).  Those radio lobes have a much larger, fainter extensions (Taylor et
al.\ 2002, in preparation), one of which is just inside the cold front
(Fig.\ \ref{fig:hydrad}).  Such coincidences suggest causal relation between
the gas sloshing and the bubbles (although, of course, it may simply be a
coincidence --- in fact, the southern radio lobe in Hydra A is bent in the
presumed direction of the gas sloshing, as if the lobes are simply entrained
by the moving gas).  One can speculate that when the bubbles rise, they may
induce bulk motion of the surrounding gas, as is indeed seen in some
simulations (e.g., Quilis et al.\ 2001).  More detailed simulations are
needed to study the viability of this hypothesis.

\section{Thermal conduction}
\label{sec:cond}

\subsection{Across cold fronts}

\begin{figure}
\vspace{5mm}
\centering \leavevmode \epsfxsize=7cm 
\epsfbox{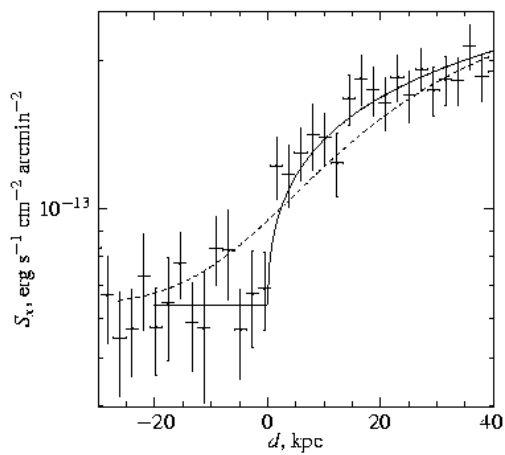}
\caption{X-ray brightness profile of the cold front in A3667. Dashed line
  shows a model of an abrupt jump convolved with a 20 kpc Gaussian that
  corresponds to the classical electron mean free path.  The front is
  significantly sharper, indicating that transport processes across the
  front are suppressed (from Vikhlinin et al.\ 2001b).}
\label{fig:a3667width}
\end{figure}

Ettori \& Fabian (2000) pointed out that the observed temperature jumps in
A2142 require that thermal conduction across cold fronts be suppressed by a
factor of 100 or more, compared to the classical Spitzer or saturated values.
Although their argument does not take into account the possibility that as
the subclusters move, the heated boundary layer may be continually stripped
and the temperature gradient thus continually sharpened, their conclusion
should be qualitatively correct.  As shown by Vikhlinin et al.\ (2001b) for
A3667, the gas density discontinuity at the cold front is very narrow --- in
fact, several times narrower than the electron mean free path, as shown in
Fig.\ \ref{fig:a3667width}.  (In some other clusters, e.g., A2142, the front
width is also unresolved by \chandra, but the data quality does not allow
such accurate constraints.)  This means that diffusion, and, by inference,
thermal conduction, are suppressed across the front, most likely by a
magnetic field parallel to the front, the same field configuration that
suppresses the growth of instabilities
(\S 2) and prevents gas stripping.

\begin{figure}[b]
\centering \leavevmode \epsfxsize=10cm 
\epsfbox{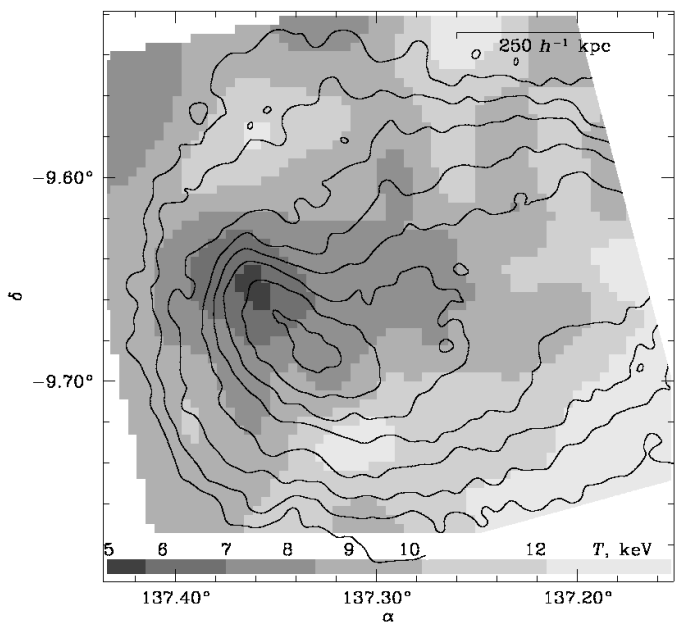}
\vspace{-4mm}
\caption{\chandra\ temperature map of a merging cluster A754 ($z$=0.05,
  $T$=9 keV), with X-ray brightness contours overlaid.  Small-scale
  temperature variations give an upper limit on thermal conduction in the
  gas.}
\label{fig:a754}
\end{figure}

Another place where thermal conduction was estimated is the boundary between
the hot cluster gas and the dense cool gas clouds observed by \chandra\ in
the centers of two cD galaxies in Coma (Vikhlinin et al.\ 2001c).  For these
clouds to survive evaporation, conduction at their outer boundaries has to
be strongly suppressed (presumably due to the disjoint magnetic field
structures).  Inside the clouds, conduction of order the classical value is
required to maintain the observed temperature gradient.

\subsection{Conduction in the bulk of the gas}

Both measurements mentioned above correspond to very special places in
clusters --- the boundaries of different gas phases that are likely to have
disjoint magnetic fields. They do not give the conduction in the bulk of the
gas, which has attracted renewed attention, e.g., due to the recent works
proposing that heat transport from the outer cluster regions into the cool
center can offset radiative cooling if the conduction is above 1/10 of the
Spitzer value (Fabian, Voigt, \& Morris 2002c; also references therein).

Conduction in the bulk of the gas can be estimated from the temperature maps
of merging clusters, such as A754 shown in Fig.\ \ref{fig:a754} (Markevitch
et al., in preparation).  In the absence of obvious strong shocks, the
merger Mach number is near 1, which gives the timescale of a merger, $t_{\rm
age} \sim L/c_s \approx 7\times 10^8$ yr, where $L$ is the size of the
merger region and $c_s$ is the sound speed.  The temperature map in Fig.\
\ref{fig:a754} exhibits spatial variations on linear scales starting from
$l\sim 100$ kpc and greater.  Assuming Spitzer conductivity $\kappa$,
these small-scale nonuniformities should disappear on a timescale $t_{\rm
cond} \sim k\, n_e\, l^2 / \kappa \approx 0.4-1 \times 10^8$ yr.  However,
they are still observed throughout the merger region and therefore should
have survived for the time $\sim t_{\rm age}$.  This requires that
conduction be suppressed by a factor $t_{\rm age}/t_{\rm cond} \sim 10$ from
the Spitzer value.

The published \chandra\ temperature map of A2163 (Markevitch \& Vikhlinin
2001) also gives a similar estimate of the conductivity; the recent long
reobservation of A2163 will provide a more accurate constraint.  In the
future, a more direct measurement may come from temperature profiles across
shock fronts, such as that in 1E0657.

\section{Relativistic electrons and thermal gas}

At radio frequencies, some clusters exhibit large, low surface brightness
halos with relatively steep spectra (for recent work see, e.g., Giovannini,
Tordi, \& Feretti 1999 and these proceedings).  They are generated by a
population of ultra-relativistic electrons that emit synchrotron radiation
in the cluster magnetic field.  The source of such electrons that can
support a cluster-size halo for a sufficient time is still unclear; several
possibilities were proposed (e.g., Harris et al.\ 1980; Tribble 1993; Dolag
\& En{\ss}lin 2000 and references therein) and extensively discussed at this
conference (see, e.g., contributions by Blasi; En{\ss}lin; Jones).  A
slightly favored view is reacceleration by high-$M$ merger shocks of
moderately relativistic electrons that were either accelerated in the past by
merger/infall shocks and/or turbulence, or produced by the interaction of
cosmic ray protons with the ICM protons.

\begin{figure}
\plotone{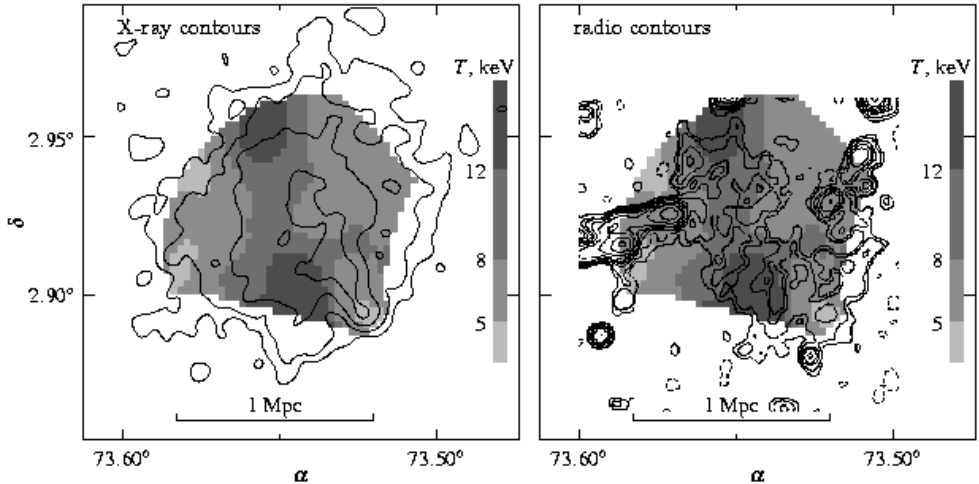}
\vspace{-7mm}
\caption{\chandra\ temperature map of a merging cluster A520 ($z$=0.2,
  $T$=10 keV). Overlaid are X-ray contours ({\em left}) and radio halo
  contours ({\em right}, from Govoni et al.\ 2001). The radio brightness
  correlates with the gas temperature.}
\label{fig:a520}
\end{figure}

\chandra\ temperature maps of clusters with radio halos can provide useful
data for this discussion.  Maps of the mergers A2163 and A665 (Markevitch \&
Vikhlinin 2001) and 1E0657 (Markevitch et al.\ 2002) showed a correlation
(although somewhat ambiguous) between the temperature and the radio surface
brightness, suggesting that relativistic electrons are accelerated in merger
shocks.  A more convincing example of such a correlation is A520 whose
temperature map is shown in Fig.\ \ref{fig:a520} along with the radio
contours from Govoni et al.\ (2001).  The X-ray data indicate a merger of a
small dense subcluster and a more diffuse main cluster.  The subcluster
appears to have left a trail of shocked hot gas, and the radio halo emission
comes predominantly from that hot region.

According to the current view, shocks with $M>3-4$ are needed to accelerate
the radio halo electrons (see, e.g., these proceedings).  It is unlikely
that in either of the above clusters the Mach number of the merger reaches
such high values (except maybe in 1E0657), so these correlations are
difficult to explain.  However, more accurate X-ray and radio measurements
for a greater cluster sample are needed to draw conclusions on acceleration
mechanisms.

\section{Summary}

Shock fronts and cold fronts discovered by \chandra\ provide unique tools to
study the physics of the ICM. In particular,

1. In merging clusters, we can estimate the velocity of the infalling
subclusters.  The cold fronts are too sharp and regular for their velocity,
indicating that gas dynamic instabilities at the boundaries between the gas
phases are often suppressed, most probably by specially structured magnetic
fields.

2. Cold fronts are also observed in the cores of most ``relaxed'' clusters,
showing that the dense, cool central gas in such clusters is constantly
sloshing in the central potential dip. This has several interesting
implications: (a) the X-ray hydrostatic method should systematically
underestimate the total cluster mass inside $r\lax 100$ kpc; (b) the
dissipated kinetic energy of the sloshing gas is sufficient to compensate
for radiative cooling and thus disrupt cooling flows.  Typically, such
sloshing involves only a small fraction of the total gas mass, so these
findings cannot be generalized to the whole cluster.

3. Thermal conduction appears to be suppressed, compared to the Spitzer
value: (a) across cold fronts, completely, which follows from the sharpness
of the gas density jumps; (b) in the bulk of the gas, by a factor $\sim 10$,
from the existence of small-scale temperature variations in mergers.

Also, we find that relativistic electrons appear to concentrate in the
recently shock-heated gas, even though in most clusters, the merger Mach
numbers are apparently insufficient to accelerate electrons to the radio
halo energies.



\begin{references}

\reference{} Allen, S.~W., Schmitt, R. W., \& Fabian, A.~C.\ 2002,
astro-ph/0205007

\reference{} Bialek, J., Evrard, A., \& Mohr, J. 2002, astro-ph/0207183

\reference{} B{\"o}hringer, H., Voges, W., Fabian, A.~C., Edge, A.~C., \&
Neumann, D.~M.\ 1993, MNRAS, 264, L25

\reference{} B{\"o}hringer, H., Matsushita, K., Churazov, E., Ikebe, Y., \&
Chen, Y.\ 2002, A\&A, 382, 804

\reference{} Buote, D. A. \& Tsai, J. C. 1996, ApJ, 458, 27 

\reference{} Churazov, E., Forman, W., Jones, C., \& B{\"o}hringer, H.\
2000, A\&A, 356, 788

\reference{} Churazov, E., Br{\"u}ggen, M., Kaiser, C.~R., B{\"o}hringer,
H., \& Forman, W.\ 2001, ApJ, 554, 261

\reference{} Churazov, E., Sunyaev, R., Forman, W., \& B{\"o}hringer, H.\
2002, MNRAS, 332, 729

\reference{} David, L.~P., Nulsen, P.~E.~J., McNamara, B.~R., Forman, W.,
Jones, C., Ponman, T., Robertson, B., \& Wise, M.\ 2001, ApJ, 557, 546

\reference{} Dolag, K., \& En{\ss}lin, T. A. 2000, A\&A, 362, 151

\reference{} Ettori, S., \& Fabian, A. 2000, MNRAS, 317, 57L

\reference{} Fabian, A.~C., Allen, S.~W., Crawford, C.~S., Johnstone, R.~M.,
Morris, R.~G., Sanders, J.~S., \& Schmidt, R.~W.\ 2002a, MNRAS, 332, L50

\reference{} Fabian, A.~C., Celotti, A., Blundell, K.~M., Kassim, N.~E., \&
Perley, R.~A.\ 2002b, MNRAS, 331, 369

\reference{} Fabian, A.~C., Sanders, J.~S., Ettori, S., Taylor, G.~B.,
Allen, S.~W., Crawford, C.~S., Iwasawa, K., \& Johnstone, R.~M. 2001, MNRAS,
321, L33

\reference{} Fabian, A.C., Voigt, L. M., \& Morris, R. G. 2002c,
astro-ph/0206437

\reference{} Giovannini, G., Tordi, M., \& Feretti, L. 1999, New Ast., 4, 141

\reference{} Govoni, F., Feretti, L., Giovannini, G., B{\" o}hringer, H.,
Reiprich, T.~H., \& Murgia, M.\ 2001, A\&A, 376, 803

\reference{} Harris, D. E., Kapahi, V. K., \& Ekers, R. D. 1980, A\&AS, 39,
215 

\reference{} Kempner, J., Sarazin, C., \& Ricker, P.\ 2002, astro-ph/0207251

\reference{} Markevitch, M., et al.\ 2000, ApJ, 541, 542

\reference{} Markevitch, M., \& Vikhlinin, A.\ 2001, ApJ, 563, 95

\reference{} Markevitch, M., Vikhlinin, A., \& Mazzotta, P.\ 2001, ApJ,
562, L153

\reference{} Markevitch, M., Gonzalez, A.~H., David, L., Vikhlinin, A.,
Murray, S., Forman, W., Jones, C., \& Tucker, W.\ 2002, ApJ, 567, L27

\reference{} Mazzotta, P., Markevitch, M., Vikhlinin, A., Forman, W.~R.,
David, L.~P., \& VanSpeybroeck, L.\ 2001, ApJ, 555, 205

\reference{} McNamara, B.~R.~et al.\ 2000, ApJ, 534, L135 

\reference{} McNamara, B. R., 2002, astro-ph/0202199

\reference{} Miralda-Escud\'e, J., \& Babul, A. 1995, ApJ, 449, 18

\reference{} Nagai, D., \& Kravtsov, A. 2002, astro-ph/0206469

\reference{} Nulsen, P.~E.~J., David, L.~P., McNamara, B.~R., Jones, C.,
Forman, W.~R., \& Wise, M.\ 2002, ApJ, 568, 163

\reference{} Peres, C. B., Fabian, A. C., Edge, A. C., Allen, S. W.,
Johnstone, R. M., \& White, D. A. 1998, MNRAS, 298, 416

\reference{} Peterson, J.~R.~et al.\ 2001, A\&A, 365, L104

\reference{} Quilis, V., Bower, R.~G., \& Balogh, M.~L.\ 2001, MNRAS, 328,
1091

\reference{} Sun, M., Murray, S.~S., Markevitch, M., \& Vikhlinin, A.\ 2002,
ApJ, 565, 867

\reference{} Tribble, P. 1993, MNRAS, 263, 31

\reference{} Vikhlinin, A., \& Markevitch, M. 2002, Astronomy Letters, 28, 495

\reference{} Vikhlinin, A., Markevitch, M., \& Murray, S.~S.\ 2001a, ApJ,
549, L47

\reference{} Vikhlinin, A., Markevitch, M., \& Murray, S.~S.\ 2001b, ApJ,
551, 160

\reference{} Vikhlinin, A., Markevitch, M., Forman, W., \& Jones, C.\ 2001c,
ApJ, 555, L87

\end{references}
\end{document}